# EVOLUTION OF FAINT IRAS GALAXIES[*]

SEB OLIVER

*Astrophysics Group, Imperial College, London. SW7 2BZ, U.K.*
*s.oliver@ic.ac.uk*

TOM BROADHURST, MICHAEL ROWAN-ROBINSON, WILL SAUNDERS,
ANDY LAWRENCE, RICHARD MCMAHON, CAROL LONSDALE,
PERRY HACKING, ANDY TAYLOR, TIM CONROW

## ABSTRACT

There has been some disagreement about the strength of evolution exhibited by IRAS galaxies. With a parameterisation such that the co-moving density increases as $(1+z)^P$ measurements of $P$ range from $2\pm 3$ to $6.7\pm 2.3$. We have recently completed a deep IRAS redshift survey which should help to clarify the situation. Paying particular attention to possible systematic effects (notably Malmquist Bias) we claim positive detection of evolution within this sample at the $3\sigma$ level. Our best estimate of the strength of evolution seen in this survey is $P = 5.6 \pm 1.6 \pm 0.7$ corresponding to luminosity evolution $L(z) = L_{z=0}(1+z)^{3.3\pm 0.8 \pm 0.6}$. This estimate is consistent with all previous determinations of the rate of cosmological evolution of IRAS galaxies.

Neither this survey nor any previous survey is sufficiently deep to distinguish between pure luminosity and pure density evolution nor between different parameterisation of these forms.

## 1 Introduction

The first indication that the IRAS galaxy population was strongly evolving came from $60\mu m$ source-counts.[1-9] An extra dimension is added with redshift information and the analysis of a collection of IRAS galaxy redshift samples (principally the QDOT sample[10]) required a strongly evolving luminosity function, consistent with the source counts data.[11] Using a parameterisation such that the co-moving density increases as $(1+z)^P$ it was claimed that $P = 6.7 \pm 2.3$, while with a parameterisation such that the luminosity increases exponentially with lookback time $L(z) = L_{z=0} \exp(Q\tau(z))$ they found $Q = 3.2 \pm 1$. This sample was insufficiently deep to distinguish between luminosity and density evolution, nor to constrain the mathematical form. In contrast, the 1.2 Jy survey was apparently consistent with no evolution of the luminosity function; $P = 2 \pm 3$.[12]

While these results are formally consistent, we are presented with two rather conflicting conclusions: the first that the IRAS galaxies are evolving as rapidly as quasars (N.B. such strong evolution could easily come into conflict with the observed

---

FIR background[8]); the second that there is little evidence for evolution. We undertook a new redshift survey of faint IRAS sources primarily to clarify the strength of evolution in these galaxies.

The analysis of the evolution of this sample will be described in greater detail in a forthcoming paper.[13]

## 2 The IRAS sample

We have undertaken a deep redshift survey of IRAS galaxies, paying careful attention to the sample selection and exhaustive follow up to ensure the maximum possible completeness and reliability.

The source catalogue was the FSDB.[14] We selected three areas near the North Galactic Pole which are relatively free from Galactic cirrus contamination (as seen in 100 $\mu m$ maps[15]) that also had good IRAS coverage at $60\mu m$. These areas thus have IRAS $60\mu m$ data with low noise so we could construct a sample with $S_{60} \geq 0.2$Jy which is highly complete. Modeling the noise distribution of our sample suggests that the thresholds applied in constructing the FSDB cause $< 1\%$ incompleteness.[16]

The total sample area is 871 square degrees within which we have 2000 extragalactic sources and 1816 redshifts. In the following analysis we exclude sectors with worse than usual redshift completeness arriving at a sample of 1437 galaxies for which we have 96% of the redshifts. We expect the incompleteness to be more significant at higher redshifts so we have also excluded sources with $z > 0.3$ and since large-scale structure dominates at low redshifts we have excluded all sources with $z < 0.02$. The final sample comprises 1207 galaxies.

We have a second survey of roughly similar size which is intended for large-scale structure studies. This sample suffers more from incompleteness and we have not yet included this in any evolutionary analysis.

Further details of these surveys are available[17] and will be published in the near future.[18]

## 3 Flux Errors

A particularly dangerous bias for evolutionary studies of flux limited surveys is the 'Malmquist Bias' (see Section 5.2). Correcting for this effect requires a detailed understanding of the flux errors in the survey. Flux errors can be divided into those that scale with source flux (fractional errors), and those which are constant at all fluxes. The first of these make a negligible contribution to 'Malmquist bias' while the constant errors are much more significant.

In the construction of the FSS two independent signal-to-noise estimates were derived:[14] the first from the scatter between successive scans of the same source (LSNR) and the second from the width of the flux/pixel distribution over a large area in the sky maps, assuming that all fluxes are simply noise (SNR). The first of these estimates will not include any component due to confusion noise. We estimate the confusion noise to be 16 mJy from an extrapolation of the source-counts (assuming negligible cirrus confusion), allowing us to compare the LSNR and SNR. The mean LSNR is well modeled by the empirical relation $S_{60}/\text{LSNR} = (0.11 S_{60} + 0.1\text{Jy})/\sqrt{\text{NOBS}}$ (where NOBS is the number of observations of the source). Using $\langle \text{NOBS} \rangle = 13.8$ and adding 16 mJy confusion noise in quadrature we find the noise at the flux limit is 36.5 mJy. The SNR gives us an independent estimate of the average noise with $\langle S_{60}/\text{SNR} \rangle = 35.8$ mJy, in very good agreement with the estimate from the LSNR. Both of these estimates include fractional errors, which we know do not significantly contribute to Malmquist bias, however we can use these as a conservative upper limit of 36 mJy to the constant error. If one were to trust the empirical LSNR model we could set the term proportional to flux to zero, giving us an estimate for the constant error of 31 mJy.

The flux scale of the IRAS PSC is quite possibly non-linear, this has the effect of mimicing evolution. We have made independent tests on the linearity of the FSS scale and believe that there is no significant problem.

The flux scale and flux errors of our survey will be discussed in detail at a later date.[16]

## 4  QDOT revisions

Since the analysis of S90 there have been a number of revisions to the QDOT data.[10] These have in general conspired to reduce the implied evolutionary strength. The corrections for a systematic redshift error in particular reduce the rate by $\delta P \sim -0.5$. Revised K-corrections also reduce the evolutionary rate, ($\delta P \sim -1$). Flux errors in the PSC have been estimated as $\sim 0.05$ Jy from comparisons of observed stellar colours with model expectations.[19] The resulting Malmquist bias will further reduce the implied evolution ($\delta P \sim -0.5$). Finally, non-linearities in the PSC flux scale may also mimic evolution ($\delta P \sim -0.5$). So, in conclusion, a 'best bet' evolutionary rate for QDOT would be $P = (4.2) \pm 2.3$.

## 5  Methods

In order to test for evolution and to test the strength of that evolution we have applied a number of standard tests and one new one.

Possibly the most basic test for evolution of an extra-galactic population is $\langle V/V_{\max} \rangle$.[20,21] Where $V$ is the available volume interior to each object and $V_{\max}$ is the available volume interior to the maximum distance at which that object would have been visible. For a non-evolving sample $\langle V/V_{\max} \rangle = 1/2 \pm 1/\sqrt{(12N)}$.

Variants of the $\langle V/V_{\max} \rangle$ test give us a measure of the strength of evolution. With the first of these one modifies the volumes to take into account any evolutionary model, the parameters of this model are varied until the $\langle V'/V'_{\max} \rangle$ is consistent with the canonical value of $1/2$.[22] A related modification is to calculate the $\langle V'/V'_{\max}\rangle$ in luminosity bins and find the best fit evolutionary parameters by minimising the $\chi^2$ over these bins assuming an expectation of $1/2$. A final $\langle V'/V'_{\max}\rangle$ variant is the maximum likelihood technique.[11,12]

A new technique has also been developed which allows one to compare a nearby survey with a more distant one directly.[13] The two important observations we have for each galaxy are their flux and distance, thus an obvious quantity to model is the number of galaxies $dN$ in a given flux and redshift bin $dSdz$. In general that distribution is given by

$$dN(S,z) = \sum_i \int_{V(z-dz/2)}^{V(z+dz/2)} \int_{L(S-ds/2,z)}^{L(S+dS/2,z)} \frac{\Omega_i(L,z)}{4\pi} \Phi_i(L) d\log L \, dV \qquad (1)$$

where $i$ represents each class of galaxy present; $\Omega_i(L,z)$, is the solid angle within which a galaxy of type $i$, luminosity $L$ and redshift $z$ would be visible; and $\Phi_i(L)$ is the luminosity function of that class. The different classes allow a multiple of $K$-corrections to be applied. We use a local sample (in this case the 2Jy Survey) using each individual galaxy as a particular 'class'. The $K$-correction for this 'class' is determined from its observed FIR colours while the luminosity function for that 'class' is taken to be $1/V_{\max}$. Flux errors are straightforwardly incorporated in the predicted flux distribution. Evolution is then applied so as to minimise the scatter between the observed and predicted dN(S,z). The results from this final method will be presented elsewhere.[13]

5.1 *K-corrections*

It has been claimed that the maximum likelihood $\langle V/V_{\max}\rangle$ method requires that the K-corrections be known to second order in redshift i.e. that the SED is known to second order in frequency.[12] We have made a number of simple models for the SED in order to test the sensitivity of our analysis to the K-corrections.

- Power law spectral index: $S(\nu) \propto \nu^\alpha$, $\alpha = -1$,

- Power law spectral index: $\alpha = -2$,

- Power law spectral index: $\alpha_i$ determined from fit to available fluxes

- Fisher *et al.* mean SED.[12]

- Two component model SED (starburst[23] and disk[24]) with relative contributions calculated using $S_{100}/S_{60}$.

The third model has the advantage that the K-correction of warm and cool galaxies will be different. The fourth K-correction is defined to second order in redshift, but does not vary from object to object. The final model has both of these features and is probably the best available to us given that the majority of our objects only have 60 and 100 micron fluxes. We decided not to apply a specific correction to the band pass (although the Fisher *et al.* SED has this built in) because the correction at 60 micron for a spectral index in the range $\alpha = -1$ to $\alpha = -2.5$ is less than 1% (Explanatory Supplement Table VI.C.6).[25]

Applying each of these methods to the $\langle V/V_{\max}\rangle$, weighted $\langle V/V_{\max}\rangle$ and maximum likelihood $\langle V/V_{\max}\rangle$ techniques gave very similar results, with the exception of the $\alpha = -2$ model which gave moderately higher rates of evolution. We thus applied the two component model in all further tests.

## 5.2 Malmquist Bias

For a population whose differential source counts decrease with flux the presence of *constant* flux errors cause more sources to be scattered above a flux limit than are scattered below it. This 'Malmquist Bias' artificially increases the slope of the source counts at the faint end or, equivalently, causes an overestimate of the faintest fluxes, directly mimicing evolution.

The effect of this bias on the differential counts for an underlying Euclidian count slope (similar to the IRAS counts) and Gaussian flux errors has been tabulated.[26] A crude parametric fit to their results is

$$R = \frac{(dN/dS)_{\text{true}}}{(dN/dS)_{\text{obs}}} \approx 1 + \frac{3.95}{(S/\sigma)^2} + \frac{56}{(S/\sigma)^4} \quad (2)$$

and

$$\langle S_{\text{true}}\rangle \approx S_{\text{obs}} R^{-0.4} \quad (3)$$

As discussed in Section 3 the constant error (which is the only significant contributor to Malmquist bias) is probably 31 mJy and not more than 36 mJy, for this analysis we conservatively assume that the errors are 36 mJy.

## 6 Large-Scale Structure

It is unfortunately impossible to distinguish evolutionary effects from large-scale structures. The answer is to sample a volume sufficiently large that it constitutes a 'fair sample' at all relevant redshifts. For this reason we excluded galaxies closer than $z = 0.02$ where survey volume would be particularly small.

One survey area was intentionally well separated from the other two allowing us to asses the contribution of the LSS in addition to reducing its impact. We have determined the evolutionary strengths for these areas independently.

Table 1 shows that there are differences between the separate areas. Interestingly these differences are less pronounced in the maximum likelihood analysis, perhaps suggesting that it is less sensitive to LSS. This demonstrates the possible dangers in drawing conclusions about evolution from any study of a single area on the sky.

Table 1: Evolutionary determinations from separate survey areas

| Areas | $N_{\mathrm{gals}}$ | $\langle V/V_{\mathrm{max}}\rangle$ | $P_{\langle V'/V'_{\mathrm{max}}\rangle}$ | $P_{\mathcal{L}}$ |
|---|---|---|---|---|
| A & E | 328 | $0.552 \pm 0.016$ | $11.6 \pm 3.8$ | $8.4 \pm 3.0$ |
| N | 879 | $0.516 \pm 0.010$ | $3.4 \pm 2.0$ | $5.5 \pm 1.7$ |

Table 2: Quantification of evolutionary strength

| Method | $P$ | $Q$ | $\chi^2(\nu = 5)$ |
|---|---|---|---|
| $\langle V'/V'_{\mathrm{max}}\rangle$ | $5.49 \pm 1.8$ | $3.74 \pm 0.89$ | |
| Binned $\langle V'/V'_{\mathrm{max}}\rangle$ | $4.94 \pm 1.4$ | $2.61 \pm 0.78$ | 11.3, 11.1 |
| Max $\mathcal{L}$ | $6.25 \pm 1.5$ | $3.54 \pm 0.71$ | |

Our final analysis uses both these regions and any LSS features in one should not be correlated with those in the others, so the impact of LSS on our conclusions is reduced. It will be possible to use these separate survey areas to asses any remaining systematic error arising from the LSS.

## 7 Results

The $\langle V/V_{\mathrm{max}}\rangle$ for this survey came out as $0.526 \pm 0.008$ i.e. a $3\sigma$ detection of evolution.

We applied two different models for the evolution. The first being pure density evolution going as $(1 + z)^P$, the second being luminosity evolution going as $(1 + z)^Q$. The values of these parameters obtained by our various methods are provided in Table 2.

The high $\chi^2$ in the binned $\langle V'/V'_{\mathrm{max}}\rangle$ method may well be unconnected to evolutionary effects e.g. incompleteness at high luminosities or large-scale structure. Alternatively it may be due to the naivity of our models which assume the same evolution for all objects regardless of luminosity and consider only the two extremes of pure density and pure luminosity evolution.

Reassuringly all methods agree within their statistical errors. By taking the average of the rates of evolution we can minimise any systematic errors arising from the method applied and assess these errors. This suggests that $P = 5.6 \pm 1.6 \pm 0.7$ where the second error indicates the spread arising from the different methods, similarly luminosity evolution would demand $Q = 3.3 \pm 0.8 \pm 0.6$.

## 8 Conclusion

We have analysed a new faint IRAS galaxy survey with the intention of establishing the true strength (if any) of this populations evolution. We have paid particular at-tention to possible systematic biases, notably the Malmquist bias and $K$-corrections. We detect evolution at the $3\sigma$ level. Our minimum estimate of the strength of this evolution would be an increase in space density of $(1+z)^{5.5\pm1.6\pm0.7}$ or an increase in luminosity $(1+z)^{3.3\pm0.8\pm0.6}$. These estimates do not yet include an estimate of the systematic errors arising from the finite volume of our survey. While it is perhaps slightly dangerous to compare evolution from different flux limited surveys which will have different population mixes, we note that our results are consistent with those from the 1.2 Jy survey and the latest analysis of QDOT.

SJO and ANT acknowledge receipt of an SERC student grant while undertaking much of this work. WS acknowledges an SERC advanced fellowship. We also gratefully acknowledge the use of the ING telescopes where the redshift survey was undertaken. We would like to thank the organisers for a very enjoyable conference.